\documentclass{svjour3}                     
\smartqed  
\usepackage{graphicx}
\begin{document}

\title{Estimating the parameters of the Sgr A$^*$ black hole}


\author{F. De Paolis         \and
        G. Ingrosso \and A.A. Nucita \and A. Qadir \and A.F.
        Zakharov
}


\institute{F. De Paolis and G. Ingrosso \at
              Dipartimento di Fisica, Universit\`a del Salento,
              Via Arnesano, I-73100 Lecce, Italy  and \\
              INFN, Sezione di Lecce, I-73100 Lecce, Italy \\
              Tel.: +39-0832-297493\\
              Fax: +39-0832-297505\\
              \email{francesco.depaolis@le.infn.it}  \\
              \email{ingrosso@le.infn.it}         
           \and
           A.A. Nucita \at
              Centre d'Etude Spatiale des Rayonnements,
              UMR 5187, av du Colonel Roche BP 44346, 31028 Toulouse Cedex 4,
              \email{achille.nucita@le.infn.it}
              \and
              A. Qadir \at
              Center for Advanced Mathematics and Physics, National
University of Science and Technology, Campus of College of
Electrical and Mechanical Engineering, Peshawar Road, Rawalpindi,
Pakistan,
\email{aqadirmath@yahoo.com}
\and A.F. Zakharov \at Institute of Theoretical and
Experimental Physics, B. Cheremushkinskaya 25, 117259 Moscow,
Russia and  Bogoliubov Laboratory of Theoretical Physics, Joint
Institute for Nuclear Research, 141980 Dubna, Russia, \email{zakharov@itep.ru}}

\date{Received: date / Accepted: date}

\maketitle

\begin{abstract}
The measurement of relativistic effects around the galactic center
may allow in the near future to strongly constrain the parameters
of the supermassive black hole likely present at the galactic
center (Sgr A$^*$). As a by-product of these measurements it would
be possible to severely constrain, in addition, also the
parameters of the  mass-density distributions of both the
innermost star cluster and the dark matter clump around the
galactic center.

\end{abstract}

\section{Introduction}
J. A. Wheeler was one of the greatest scientists of the
twentieth century because of his many contributions to nuclear
physics, theory of quantum measurement, general relativity and
relativistic astrophysics. He had a talent to express complex
concepts in simple and catchy form. In particular, he coined
the term ``black hole'' which has gained enormous popularity
even among the non-scientists.  He has been called the greatest
un-crowned Nobel laureate. His way of writing was very
attractive for readers because he was a great scientist,
teacher and writer in one personality. The last scientific
paper by Wheeler was on retro-MACHOs (we prefer to call them
retro-lensing or relativistic images), written in collaboration
with D. Holz \cite{hw}. The authors took seriously something
that was well known to be possible around a black hole (BH) in
principle. They considered our Sun as the source of light rays
that, if the impact parameter is right, may go around the BH
and reach the observer forming a ring around the BH. Light rays
with a slightly smaller impact parameter may go twice around
the BH and then escape to the observer forming a slightly
smaller ring, and so on. So, in the case of perfect alignment,
an infinite series of rings should appear and therefore they
suggested that a  survey of the sky be made to look for these
rings as a way to search for isolated BHs near the Sun.

In general, Earth moves on its orbit so that the retro-lensing
image magnitude changes with time in a periodic way. The problem
is that our Sun is not a very bright source and even using the
HST, only a BH heavier than 10 $M_{\odot}$ within 0.01 pc might be
revealed in this way. Although it may be expected that sooner or
later a survey of the sky to search for such rings will take
place, it is for the moment much better to consider binary systems
composed of massive BHs and luminous stars and look, therefore,
towards well known BH candidates. This is exactly what has been
suggested in \cite{s2} soon after the Holz and Wheeler paper
appeared, that is the supermassive black hole at the galactic
center (Sgr $A^*$) is the most interesting retro-lens to look
for\footnote{Similar results had already been pointed out in
\cite{ve} (see also \cite{virbhadra} as a recent paper on this
issue).}.

A classical method for estimating the physical parameters (in
particular mass and angular momentum) of BHs and in particular of
the candidate supermassive black hole at Sgr $A^*$ is to look for
the periastron or apoastron shift of the stars orbiting around it
(the so-called S-stars). In Section 2 we shall briefly review this
issue. However, the amount of the apoastron shift strongly depends
also on the distribution of both the stars and the dark matter
making it practically impossible to estimate the BH parameters.

A more direct way to estimate the central BH parameters,
including in principle also its electric charge) is to measure
the shape of the retro-lensing images of the brightest and
innermost stars in the K-band of the electromagnetic spectrum
(or also the shape of the BH shadow in the radio band). This
issue, together with the information that can be obtained by
measuring the spectrum (effect not considered in detail in the
literature) of the stellar retro-lensing images will be
discussed in Sections 3 and 4.

By combining these measurements, also taking into account the
proper motion of the supermassive BH, it will be possible in
the near future to estimate both the BH parameters and those of
the stellar and dark matter distributions.

\section{S-stars apoastron advance: a laboratory to measure the
supermassive BH parameters}

The precise measurements of the velocity dispersion of the S-stars
orbiting around  Sgr$A^*$ have allowed us, in the last decade, to
constrain more and more strongly the mass enclosed in a smaller
and smaller region around the galactic center. With the Keck 10 m
telescope, the proper motion of several stars orbiting the
Galactic Center have been accurately monitored and the entire
orbit of the S2 star has been measured allowing an unprecedented
description of the Galactic Center region (for a review see
\cite{alexander2005}). The estimates of the supermassive BH mass,
or more exactly of the amount of mass M($< r$) contained within a
certain distance $r$ from the galactic center, changed
dramatically during the last decade from about $2.6\times
10^6~M_{\odot}$ within about $10^{-2}$ pc \cite{schoedel},  to
$\simeq 4 \times 10^6~M_{\odot}$ within $10^{-2}$ pc
\cite{reid2007}. Ghez et al. \cite{ghez2003,ghez2005} have found
that a mass of about $(3.67\pm 0.19)\times 10^6~M_{\odot}$ must be
present within the central $3\times 10^{-4}$ pc of the galactic
center.

An important point to be mentioned is that the identification of
the super massive BH at the galactic center is also based on the
observations from radio to Near IR to X-ray wavelengths
\cite{narayan}. The bolometric luminosity of Sgr A$^*$ is  $\simeq
2\times 10^{36}$ erg s$^{-1}$, that is  $\simeq 4 \times 10^{-9}$
of the Eddington luminosity (this is also known as the ``blackness
problem'' and accounting for that is an important issue to be
settled in all details). This means that Sgr A$^*$ is a very faint
source, most luminous in the sub-millimeter wavelengths (explained
naturally as synchrotron emission by thermal electrons near the
BH). Also $X$ and $\gamma$ photons are detected from the direction
of Sgr A$^*$ that, however, is not resolved at these wavelengths
\footnote{Recently, a $\simeq 20$ min variability observed by
XMM-Newton in the 2004 X-ray flare towards Sgr A$^*$ has been
interpreted as due to matter falling from the innermost stable
orbit to the Schwarzschild radius of a $3.7\times 10^6~M_{\odot}$
BH at the galactic center \cite{belanger}.}. The identification of
the supermassive BH at Sgr A$^*$ may be considered as not
completely settled as yet, although the supermassive BH option
seems to be that that more naturally explains the different
observational evidences.

Further information may be obtained, in principle, from the
measurement of the stellar periastron (or apoastron) shifts
yielding rosetta-type orbital shapes. Several authors  have
discussed the possibility of measuring the general relativistic
corrections to Newtonian orbits for the BH at Sgr$A^*$,
assuming either a Schwarzschild or a Kerr BH. For a test
particle orbiting a Schwarzschild black hole of mass $M_{\rm
BH}$, the periastron shift is given by
\begin{equation}
\Delta \phi_S \simeq \frac{6\pi G
M_{BH}}{d(1-e^2)c^2}+\frac{3(18+e^2)\pi
G^2M_{BH}^2}{2d^2(1-e^2)^2c^4}~, \label{schshift}
\end{equation}
$d$ and $e$ being the semi-major axis and eccentricity of the
test particle orbit, respectively. For a rotating black hole
with spin parameter $a=|{\bf a}|=J/GM_{\rm BH}$, in the most
favorable case of equatorial plane motion ({\bf a.v} = 0), the
shift is given by (Boyer and Price \cite{boyerprice}, but see
also \cite{bini} and references therein as a recent paper on
this issue)
\begin{equation}
\begin{array}{l}
\displaystyle{\Delta \phi_K \simeq \Delta \phi_S +\frac{8a\pi
M_{BH}^{1/2}G^{3/2}}{d^{3/2}(1-e^2)^{3/2}c^3}+\frac{3a^2\pi
G^2}{d^{2}(1-e^2)^{2}c^4}~,} \label{kershift}
\end{array}
\end{equation}
which reduces to eq. (\ref{schshift}) for $a\rightarrow 0$. The
apoastron (periastron) shifts (measured in mas/revolution) as
seen from Earth (at the assumed distance of $R_0=8$ kpc from
the galactic center) is $\Delta\phi_E*^{\pm}=d(1\pm
e)\Delta\phi /R_0$, where the sign $+$ holds for the apoastron
and the $-$ for the periastron, respectively.

As discussed in \cite{nucita2007}, notice that the differences
between the periastron shifts for the Schwarzschild and the
maximally rotating Kerr black hole is at most $0.01$ mas for
the S2 star and $0.009$ mas for the S16 star. In order to make
these measurements with the required accuracy, one needs to
measure S-stars orbits with a precision of at least $10$
$\mu$as. At present\footnote{For the case of relative
astrometry, the present uncertainty in the position measurement
of a star brighter than K=15.5 is about 0.1 mas \cite{yelda}.}
this precision is not attainable by the available instruments,
but in the near future it will certainly be reached. For
example, there is a proposal to improve the angular resolution
of VLTI with the PRIMA facility (see the web site
http://obswww.unige.ch/PRIMA/home/introduction) which, by using
a phase referenced imaging technique, will get $\sim 10\mu$as
angular resolution.

However, even if at least in principle the effect of a
maximally spinning BH on the periastron shift of an S-star can
be distinguished from that induced by a Schwarzschild BH with
the same mass, actually that is, in practice, impossible due to
the presence of the stellar cluster surrounding the
supermassive BH. The effect of this stellar cluster
distribution gives (for most of the possible configurations) a
much larger effect on the periastron shift with respect to that
of the central BH. In particular, if one describes the stellar
cluster by a Plummer model with density profile
\begin{equation}
\rho_{CL}(r)=\rho_0\left[1+\left(\frac{r}{r_c}\right)^2\right]^{-\alpha/2}~, \label{pl}
\end{equation}
where the central cluster density $\rho_0$, its core radius
$r_c$ and the power slope $\alpha$ are free parameters
constrained by the condition that the total mass  $M(r)=
M_{BH}+ M_{CL}(r)$, it can be shown that if $r_c$ is smaller
than the semimajor axis of a considered star, the apoastron
shift induced by the cluster is in the same direction of that
induced by the BH, while they are in opposite directions if
$r_c$ is larger than the stellar semimajor axis (for further
details see \cite{nucita2007}). A detailed analysis shows that,
if one considers the S2 star, the transition from prograde
shift (due to the BH) to retrogade shift (due to the stellar
cluster) occurs for a cluster mass of only 0.1-0.3\% of the BH
mass for $r_c\simeq 5.5$ mpc, almost irrespectively of the
value of $\alpha$. This means that a small fraction of mass in
the cluster drastically changes the overall shift.

\section{Shadow, or gravitational retrolensing to measure the
supermassive BH hairs}

The calculation of  the retro-lensing images by numerical methods
is quite complicated because it is necessary to integrate null
geodesics with very high accuracy (see e.g. \cite{ve,ohanian,et}).
In the Schwarzschild case for example, obviously, as the impact
parameter $u$ becomes smaller and smaller the deflection angle
grows (the weak deflection limit is no longer valid at this point)
until it diverges at the limiting value of the impact parameter
$u_m=3 \sqrt3 GM/c^2$. Therefore, very large deflections are
possible for $u$ close to $u_m$. A deflection of order $\pi$ means
that the photon turns around the BH and goes back to the source
(retro-lensing as named by Holz and Wheeler). Photons with impact
parameter equal to $u_m$ are injected in a circular orbit of
radius $r_m=1.5~ r_s$ around the BH and photons with $u<u_m$ fall
inside the BH and get lost. Let's note that the angular separation
from the BH of the images formed by photons with impact parameter
u is practically given by $\theta=\arcsin(u/D_{OL})\simeq
u/D_{OL}$ and the fact that no deflection is possible for $u<u_m$
implies the existence of a lower value for the image angular
separation that is (in other words no source image can appear at
angular separation less than) $\theta_m=u_m/D_{OL}$. This minimum
angle $\theta_m$ defines the so-called shadow of the BH. The case
for spinning and/or charged black holes has been also considered
in the literature (see e.g.
\cite{carter,bozza2003,kerr,zak2005a,zak2005b,bozza2005,vazquez,bozza2006}).
The shadow of the BH is slightly larger than the BH horizon and
the shape depends on the BH parameters, as for the retrolensing
curves. Due to the strong gravitational field the light emitted
behind the BH cannot reach us  since it is deflected, trapped and
absorbed by the BH. The effect manifests itself with the absence
of radiation - the shadow -  at the boundary of the event horizon.
The dimension of the shadow should be about 30 $\mu$as. It is like
the shadow cast by the moon on the sun during a solar eclipse. The
difference in the BH case is the strong gravitational field that
makes the shadow larger than the event horizon.

The magnitude of the retrolensing images by the BH in Sgr $A^*$
has been discussed (see e.g. \cite{s2,kerr}) both for the
Schwarzschild and Kerr BH. In the Schwarzschild case, for the S2
star, they  are in the range $33.3 - 36.8$ (depending on the star
distance from the black hole) in the K-band, that turns out to be
close to the limiting magnitude of the next generation of
space-based telescopes. The BH spin effect gives only a minor
correction on the retrolensing image brightness and so it cannot
increase substantially the image amplification with respect to the
Schwarzschild black hole case \cite{kerr}. An even more important
effect induced by the BH spin is the retrolensing image
deformation \cite{zak2005a,bozza2006} since light rays co-rotating
with the BH spin can get closer to the BH horizon, while
counter-rotating photons get further away. Also the eventual BH
electric charge has an influence on the shape of the retrolensing
images in the sense that the higher the charge the smaller is the
size of the retrolensing image that becomes $r_s/2$ for a
maximally charged Reissner-Nordstrom BH \cite{zak2005b}.

The shape of the retrolensing images will be possibly measured in
the future in the K-band or even in X-rays. In the last case the
source would be an accreting neutron star or stellar mass BH
(there are many of these objects around Sgr A*, as shown by
CHANDRA). However, the first direct information about the
parameters of the BH at Sgr $A^*$ will probably come by the
detections of the shadow shape in the radio band. The present
angular resolution obtained last year by Doeleman is about 37
microarcsec at 1.3 mm \cite{doeleman}. However, at wavelengths
higher than 1 mm the scattering by electrons surrounding the BH
blurs the image making any progress in the resolution useless.
Sub-mm data would probably be able to show the shadow rather soon
and later on will be even possible to observe its shape.

The angular resolution of an instrument able to see the shape of
the shadow is close to that necessary to read a newspaper open on
the moon from Earth. In the next few years, it will be likely
possible to identify the spin of the BH in Sgr $A^*$, either by
detecting the periodic signature of hot spots at the innermost
stable circular orbit or by producing, for example by a
(sub)millimeter VLBI "Event Horizon Telescope", images of the
Galactic center emission to see the shadow silhouette. These
techniques are also applicable to the BH in M87, where the BH spin
may be key to understanding the jet-launching region.

Before closing this section we wish to emphasize that, whilemany
of the available observations in the radio, IR and X-ray bands may
or may not be interpreted as evidence for the supermassive BH at
the galactic center, the only one that would unambiguously
demonstrate the existence of such a BH is the observation of
retrolensing images towards Sgr a$^*$.

\section{Retrolens spectrum}

A novel effect - which is a consequence of the black hole spin
- on the retro-lensing image: the frequency change of the
observed light.

The effect we intend to exploit for our purposes was
essentially found by Floyd and Penrose \cite{fp}. They
demonstrated that rotational energy can be extracted from a
spinning black hole by sending in a particle co-rotating with
the hole. If the particle decays inside the ergosphere (the
region inside which particles at rest would appear to be moving
superluminally according to observers at infinity) the decay
product falling into the black hole would have a spacelike
momentum vector and so it would deposit a negative energy
according to the observer at infinity. It was later shown by
Christodoulou and Ruffini \cite{cr} that there is a limit on
the amount of energy that can be so extracted. This fact leads
one to consider what happens to a photon passing through the
ergosphere, and close to the limit it is allowed to reach
before it is pulled into the hole (the photosphere). At first
one would be tempted to answer that there is no change, as the
photon does not decay inside the ergosphere. However, this
intuitive answer takes for granted that the photon is a test
particle. One should really consider the back-reaction of the
photon on the geometry. This may be too difficult to accomplish
directly. Let us, therefore, consider the matter qualitatively.

We consider the photon as being absorbed and then emitted by
the black hole. More correctly, what has happened is that the
photon enters and then exits the region where frame dragging is
significant. This speeds up the hole by a minuscule amount if
the photon is co-rotating with it (and slows it down when it is
counter-rotating), as some angular momentum and energy are
imparted to it while it carries the photon along. Now the
photon will have somewhat more energy than when it was emitted
by the source. The longer it is ``carried along'' the more
energy is gained. The closer it gets to the hole the more the
effect is. This is simply the analogue for photons, of the
``gravitational slingshot'' used to accelerate spacecraft, that
gives us an estimate of the energy gained by the photon. In
terms of the change of frequency $\delta\nu$ we then have
\begin{equation}
\frac{\delta\nu}{\nu} \simeq \frac{Gm}{bc^2}\frac{a}{b}~,
\end{equation}
which is  ${ma}/{b^2}$ in gravitational units ($G = c = 1$),
where $m$ is the mass of the hole, $b$ the distance of closest
approach to the hole and $a$ the spin parameter of the hole
(being the angular momentum per unit mass). To be able to
obtain the numerical factor and to be surer of the result, we
need a more precise analysis.

We could also obtain the frequency change by invoking the
formula: $\delta\nu/\nu = \delta t/t$, where $\delta t$ is the
time delay due to the rotation of the hole and $t$ is the time
spent in the strong field of the hole. If the light goes round
the hole and is reflected, the time taken is the distance
travelled, which is half way round the hole (at light speed),
which is $\pi b$. The time delay is due to the cross term in
the metric, responsible for frame dragging, $2ma/b$ and acts
over an angle $2\pi$. Thus, the total effect for the spectral
shift for the $n^{\rm th}$ retro-lensed ring is
\begin{equation}
\left(\frac{\delta\nu}{\nu}\right)_n = 4 \frac{ma}{b^2}(2n-1)~.
\end{equation}
Indeed, $4ma/b^2$ is the effect that would be expected for the
primary retro-lensing image. The secondary one would have an
additional time delay of going once around, i.e. it would be 3
times this value and the next 5 times, so that for the $n^{\rm
th}$ ring it would be $(2n - 1)$ times this value.

One may worry whether the $t$ in the denominator takes on the
extra multiples and so $\delta\nu/\nu$ should be the same for
all ``rings". To check this it is only necessary to Lie
transport the momentum vector for the photon along the orbit
around the hole. It is obvious from here that the $(2n - 1)$
factor is real. Another worry could be that the endless
increase of photon energy by going around the hole many times
may seem unreasonable. To see that this is not so, notice the
analogy of the process used with the acceleration of charged
particles in an accelerator. By taking a charged particle many
times around in an appropriate electromagnetic field, one can
go on accelerating it. Here we are taking a photon around a
gravitational field and ``accelerating'' it in the sense of
increasing its energy. We may view the spinning hole as a
``giant {\it photon} synchrotron'' with very low yield.

For a counter-rotating photon, since the angle $\phi$ is
integrated in the reverse direction, the effect is negative and
we get a red shift instead of a blue shift. Naively, we might
expect a ring, blue-shifted on one side and red-shifted on the
other. This will not be the case. There are two problems with
this naive expectation. The first is that the distance of
closest approach for a Schwarzschild hole is the {\it
photosphere}, $3m/2$, but for the Kerr hole it changes. In the
co-rotating direction it coincides with the horizon for an
extreme hole and in the counter-rotating direction it is 4
times that value. As such the ring should be off the center of
the hole and the $\delta\nu/\nu$ decreased by the square of the
factor increasing $b$. Roughly speaking then, in the case of
the perfect alignment of Holz and Wheeler, we should expect -
on account of the black hole spin - a ring-like pattern one
side of which is bluer and the other redder, with the hole at
the focus closer to the bluer side. The ring will be brighter
on the bluer side and fade away from it as the redder side is
approached.

Of course, this is only the {\it Doppler} shift due to the
photon being ``accelerated" or ``decelerated". There is also
the gravitational red-shift to take into account. Since the
photon approaches closer to the source when it is blue-shifted
and is further away when it is red-shifted, the two
gravitational red-shifts are not equal. A simple computation,
for $a=1$, gives the red-shift on closer approach to be $9/29$
and on further approach to be $8/13$. The net blue-shift for
the first ring gives $\delta \nu/\nu\simeq 1.945$ and the
red-shift to be $\delta \nu/\nu\simeq -0.950$. Since the
gravitational red-shift remains the same for all the rings for
the second (and fainter) ring the total blue shift would be
$\delta \nu/\nu\simeq 7.065$ and the total red-shift would be
$\delta \nu/\nu\simeq -2.230$.

\begin{figure*}
\includegraphics[width=0.75\textwidth]{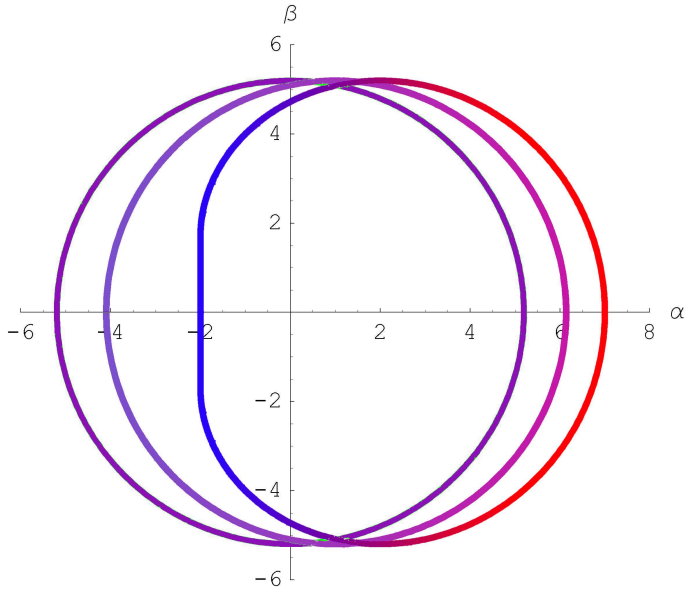}
\caption{The precise shape of the retro-lensing image in the
case of perfect alignment, with the spin perpendicular to the
line of sight, is shown. Here, $\alpha$ and $\beta$ are the
celestial coordinates of the image as seen by an observer at
infinity. The circular image corresponds to the retro-lensing
image for a Schwarzschild black hole. The second image is that
expected for a Kerr black hole with rotation parameter $a=0.5$.
The third one corresponds to a maximally rotating Kerr black
hole with $a=1$. The coloring of the rings is given to indicate
the frequency shift due to the spin effect (not giving the
precise color shifts). Note that the actual ring colors have
been simulated much more precisely for quantitative prediction
\cite{jp}. Here the black hole is assumed to rotate
counterclockwise as seen from above. The unit of length along
the coordinate axes $\alpha$ and $\beta$ is $M$, so that the
image in the Schwarzschild case is at a distance of
$3\sqrt{3}/2$ Schwarzschild radii from the black hole center.}
\label{fig}
\end{figure*}

The shape of the retro-lensing image in the perfect alignment
case is shown in Fig.~\ref{fig}. The circular shape corresponds
to a Schwarzschild black hole, the second image to a Kerr black
hole with $a=0.5$, the third one to a maximally rotating Kerr
black hole ($a=1$). Here, the black hole is assumed to rotate
counterclockwise as seen from above. As is clear from the
discussion above, in the Schwarzschild case the retro-lensing
image has everywhere the same color, since no spectral shift
effects are present in this case. For a Kerr black hole,
instead, the light rays co-rotating with the black hole form a
closer image which is also blue-shifted with respect to that
formed by counter-rotating photons. The maximal effect is
obtained for a Kerr black hole with $a=1$. In Fig.~\ref{fig},
$\alpha$ and $\beta$ are the celestial coordinates of the image
as seen by an observer at infinity which are function of  the
parameters $\xi$ and $\eta$ describing the null geodesics in
the equatorial plane (see, Chandrasekhar \cite{chandra}).

The other problem is that photons going off the equatorial plane
will not be reflected properly \cite{vazquez,bozza2006}. The
further they are off the plane the more the path is altered. As
such they will be distorted. Further, the photons for the extra
rings will be shot off at even further removed angles and so will
disappear even more sharply than would be the case for the
Schwarzschild hole. Precise simulations of the large angle
deflection images about rotating black holes have been made by
using ray tracing techniques \cite{bd,jp}. In particular,
\cite{bd} have looked at the K$_{\alpha}$ line of iron and its
Doppler shift. The inner regions of the images created in the
simulations are given in \cite{jp}

Both the above papers ignore the higher order relativistic
images discussed in the present paper. There would be a time
delay from one image to the next due to the photon going around
an extra time \cite{bm}. For Sgr A$^*$ the time delay would be
as much as $10$ minutes! The second ring, though much fainter
(on account of the lower intensity of photons), would show much
higher frequency shifts and would not be so contaminated by the
other rays that broaden the primary rings. This is because only
those rays would be able to go around the extra time that are
adequately well aligned. (In fact that is what makes the image
fainter.)

\section{Concluding remarks}

Let us now look at the possibility of really measuring in the
near future the BH parameters at Sgr A$^*$. At first let us
consider the retrolensing method and in particular the shape of
the retrolensed images. As can easily understood, this is a
more difficult task than detecting simply the retro-lensing
images. For simplicity let us confine ourselves to the
perfectly aligned case and consider only the first of the
infinite number of the retro-lensing rings. The first effect of
the black hole spin is in fact that of deforming the circular
rings expected for the Schwarzschild case. This effect is
considered in \cite{kerr}, but it is obvious that detecting the
deformation of the rings is not an easy task since the angular
resolution of the instrument must be smaller with respect to
the ring angular extent, which in turn is around $3r_s/2$. This
distance is expected to be in most cases much smaller with
respect to the angular resolution of available instruments.

As discussed in \cite{s2}, with the next generation space
telescopes like NGST \cite{ngst} it will be possible to detect in
the K-band the retro-lensing images of the S2 star orbiting around
Sgr A$^*$. However, NGST will not have the angular resolution
necessary to make the spectrum of the two sides of the image to
measure the spectral shift. Since the angular size of the
Schwarzschild radius of the Sgr A$^*$ black hole is of about
$10~\mu$as an instrument with at least that angular resolution is
necessary to eventually observe the proposed effect. In the
near-medium future, several space-based instruments may have the
necessary angular resolution such as VLBI \cite{vlbi,vlbi2} in the
radio band and MAXIM \cite{maxim} and Constellation X
\cite{constellationx} in the X-ray band, so that there is a hope
of measuring the black hole spin by simply measuring the spectrum
of a retro-lensing image. Actually, it is very likely that the
first real measurement of the central BH parameters will be made
by using radio observations detecting the shadow of the BH. Radio
observations have presently angular resolution of  $37 \mu$as at
230 GHz with a baseline of 4500 km \cite{doeleman}, that is very
close to the size of the BH shadow (about $30 \mu$as for a four
million solar mass BH). Of course, to measure the shape of the
shadow a much better angular resolution is needed, but progress in
this field is rather fast and in the next one or two decades an
angular resolution of a few $\mu$as in the radio band is
attainable. Moreover, it would be also possible in future to use
different bandwidths to determine the spectral effect discussed in
the previous section and therefore  estimate the supermassive BH
parameters.

As discussed in Section 2, the measurement of the supermassive BH
physical parameters by using  retrolensing images (or the BH
shadow), together with the measurement of the periastron (or
apoastron) shift of some of the S-stars, offers a unique
opportunity of estimating also the mass and density distribution
of the star cluster present around the BH at Sgr $A^*$. With an
instrument able to get accurate position measurements (within 10
$\mu$as, the pericenter shift measurements of only three S-stars
is enough to strongly constrain the Plummer model parameters of
the stellar cluster.

A dark matter (DM) component (for example constituted by WIMPs)
may be present around the galactic center, in addition to the
BH and the stellar cluster, and this mass component could also
modify the trajectories of the stars moving around Sgr $A^*$
significantly, depending on the DM mass distribution. One can
therefore use a three component model for the central region of
our galaxy constituted by a central BH, a stellar cluster and a
DM sphere, i.e.
\begin{equation}
M(<r)=M_{BH}+M_*(<r)+M_{DM}(<r)~.
\end{equation}
Following \cite{hallgondolo}, the DM concentration can be
described by  a mass distribution of the form
\begin{equation}
M_{DM}(<r)=\left\{
\begin{array}{ll}
M_{DM}\left(\frac{r}{R_{DM}}\right)^{3-\beta}~,~~~~~~~r\leq R_{DM} \\ \\
M_{DM},~~~~~~~~~~~~~~~~~~~~~~~r> R_{DM}
\end{array}
\right. \label{mass_dm}
\end{equation}
where $\beta$ is a free parameter and $M_{DM}$ and $R_{DM}$ are
the total amount of DM in the form of WIMPs and the radius of
the spherical mass distribution, respectively. In this case,
for the central stellar cluster, in \cite{hallgondolo} the
empirical mass profile obtained by the data has been used
\begin{equation}
M_*(<r)=\left\{
\begin{array}{ll}
M_*\left(\frac{r}{R_*}\right)^{1.6}~,~~~~~~~r\leq R_* \\ \\
M_*\left(\frac{r}{R_*}\right)^{1.0}~,~~~~~~~r> R_*
\end{array}
\right. \label{mass_star}
\end{equation}
with a total stellar mass $M_*=0.88\times 10^6$ M$_{\odot}$ and
a size $R_*=0.3878$ pc.

The present upper limit of $10$ mas on the periastron shift of
the S2 and S16 stars allows us to constrain the radius of the
dark matter distribution (assumed, following
\cite{hallgondolo}, with a total mass $M_{DM}\simeq 2\times
10^5$ M$_{\odot}$) more strongly than the results in
\cite{hallgondolo} (where it is found that there are acceptable
configurations only with size in the range $10^{-4}-1$ pc).
Consideration of the available upper limit of the stellar
periastron shift allows instead the conclusion that DM
configurations of the same mass are acceptable only for
$R_{DM}$ in the range between $10^{-3}-10^{-2}$ pc, almost
irrespectively of the DM $\beta$ value \cite{zak2007,zak2009}.

Much stronger constraints on both the DM concentration and stellar
cluster may be obtained in the future if a real measurement (and
not only an upper limit) of the periastron shift of some S-stars
will be available. With a measurement with angular resolution
about 10 $\mu$as of the periastron shift of only two stars it
would be possible to severely constrain the DM distribution if the
stellar cluster distribution is known and provided that the BH
parameters are known. With the S-stars presently known, and taking
into account their periods \cite{gillessen,bozza2009}, it would
require about 19 years of measurements of the stellar orbits with
a precision of about 10 $\mu$as to get the desired result. With
the measurement of the periastron shift of five S-stars it would
be possible to constrain the parameters of both the stellar and DM
clusters. That would require about 50 years of observations with
that precision.

\begin{acknowledgements}
One of the authors, AQ, is grateful to the INFN, Project FA51 and
Prof. I. Ciufolini for financial support at the University of
Lecce, where this work was finalized. Prof. Remo Ruffini is also
acknowledged for enlightening discussions.
\end{acknowledgements}



\end{document}